\documentstyle[12pt]{article}

\newcommand{\be}[1]{\begin{equation}\label{#1}}

\newcommand{\ee}{\end{equation}}

\newcommand{\spn}{{\rm span}}

\newcommand{\fuss}{\hbox{\rlap{\hskip2.5pt \vrule height 6pt width0.6pt}N}}
\newcommand{\supra}{\large{\$}}
\newcommand{\com}{\hbox{\rlap{\hskip2.5pt \vrule height 6pt width0.6pt}C}}
\newcommand{\real}{\hbox{\rlap{\vrule height6.5pt width0.6pt}R}}
\newcommand{\quat}{\hbox{\rlap{\vrule height6.5pt width0.6pt}H}}
\newcommand{\aker}{Z\!\!\!\! Z}
\newcommand{\copos}{\{\quat\}}
\newcommand{\qopos}{\{\supra\}}
\newcommand{\cem}{\cal{M}}
\newcommand{\qem}{{}_{q}\cem}
\newcommand{\qset}{\{{\cal{G}}\}}

\newcommand{\s}{\sigma}

\title{A Non-Classical Linear Xenomorph as a Model for Quantum Causal Space}

\author{Ioannis Raptis\footnote{Algebra and Geometry Section,
    Department of Mathematics, University of Athens,
    Panepistimioupolis 157 84, Greece; s-mail address: 4 Kondoleondos
    Street, Paleo Psychiko, Athens 154 52, Greece;
    e-mail address: iraptis@eudoxos.dm.uoa.gr}}

\date{}

\begin{document}

\maketitle

\begin{abstract}
A quantum picture of the causal structure of Minkowski space $\cem$ is
presented. The mathematical model employed to this end is a
non-classical version of the classical topos $\copos$ of real
quaternion algebras used elsewhere to organize the perceptions of
spacetime events of a Boolean observer into $\cem$. Certain key
properties of this new quantum topos are highlighted by contrast
against the corresponding ones of its classical counterpart $\copos$
modelling $\cem$ and are seen to accord with some key features of the
algebraically quantized causal set structure.  \end{abstract}

\section{INTRODUCTION}

Real four-dimensional Minkowski space $\cem$ is what a Boolean
researcher perceives (Trifonov, 1995). Her perceptions of spacetime
events, or equivalently, her states of knowledge of the world, that is
to say, her controlling actions on and `passive' observations of the
events of the world, are uniquely modelled by the topos $\copos$ of
real quaternion division algebras\footnote{A note on unusual language:
in (Trifonov, 1995) the real quaternion objects $\quat$ of this Boolean
topos are called $\real$-paradigms and the topos itself a classical
$\real$-xenomorph. The general (classical) topos theory for $\cem$ may
be coined (Classical) Linear Xenology (Trifonov in private e-mail
correspondence).} which are then seen to effectively encode in their
multiplicative structure the Lorentzian metric
$\eta_{\mu\nu}=diag(-1,+1,+1,+1)$ of $\cem$. Since from the latter
derives the causal structure at each spacetime event, the Minkowski
lightcone soldered at each event, the result above can be rephrased as
follows: a researcher that orders her perceptions of spacetime events
according to classical Boolean logic, that is, a classical researcher,
builds a unique picture of a special relativistic causal order between
them and the latter is effectively what she `sees' as the world's
chronological connection.  $\cem$ as a causal space is what a classical
researcher perceives.  This explains the reality ($\real$),
dimensionality ($4$) and signature ($+2$) characters of $\cem$ as
inevitable results of the Boolean mode of perception (logic algebras)
of spacetime events of a classical researcher and due to this $\cem$
may be called `classical causal space'.

On the other hand, it was established early (Robb, 1914, 1921,
Alexandroff, 1956 and Zeeman, 1964, 1967) and revived lately (Sorkin
{\it et al.}, 1987) that causality, when modelled by a partial order
between events, determines as well the three characteristics of $\cem$
above\footnote{The work of Sorkin {\it et al.} on a causal set theory
of the small scale structure of spacetime goes even a bit further and
argues that not only the Lorentzian metric signature, real coefficient
field and the four-dimensionality of $\cem$ are determined when a
partial order models causality between events, but also its topological
and differential structure.}.  It follows that a partial order may be
thought of as a sound model of the classical causality relation which
is compatible or in accord with the Boolean mode of perception of
spacetime events of a classical researcher. Since the latter is
organized into the classical topos $\copos$, we infer that the
multiplicative structure of quaternions is intimately related to that
of posets and both to a classical, Boolean logical perception of the
event-structure of the world. Thus, it is fairly natural to suppose
that a non-classical researcher, defined as one that employs some kind
of non-Boolean or `quantal' logic to order her perceptions of spacetime
events, builds based on it a quantum picture of the causal structure of
$\cem$, call it $\qem$.  The latter may be equivalently viewed as the
product of some sort of quantization of the classical causality
relation which is represented by a partial order in $\cem$.

In the present paper we present a candidate for such a quantum version
of the causal structure of Minkowski space based on a non-classical
model topos $\qopos$ of a new algebraic structure $\supra$ that is a
multiplicative deformation of the quaternions and was introduced in
(Raptis, 1998). We baptize the $\qopos$ model of $\qem$ `quantum
topos'\footnote{Or, to comply with (Trifonov, 1995), `quantum linear
xenomorph' and the general quantum topos theory that may be developed
from it `Quantum Linear Xenology'.} and study its properties in
comparison with those of its classical counterpart $\copos$ model of
$\cem$\footnote{This will go some, but still short, way in determining
the `true quantum topos' for quantum relativistic spacetime. The latter
may prove to be an instance of the crucial missing denominator in the
riddling analogy $\frac{\rm locales}{\rm quantales}=\frac{\rm
topoi}{\rm ?}$ that has puzzled mathematicians for a while now (Lambek
and Selesnick in private s-mail and e-mail correspondence).}.  The
$\quat$-deformed multiplicative structure of $\supra$ encodes a quantum
sort of causality distinct from the classical Minkowskian one of
$\quat$ which can be equivalently cast as a partial order as mentioned
above. A recent result from a straightforward algebraic quantization of
the classical partial order causality (Raptis, 1999) further supports
the soundness of $\qopos$ as a non-classical linear xenomorph model of
the quantum causal space $\qem$.

The paper is organized as follows: in Section 2 we recall some basic
terminology, essential facts and results from (Trifonov, 1995), mainly
that a classical researcher `uniquely determines' $4$-dimensional, real
Minkowski space as the structure of her own (proper) states of
knowledge of the events of the world, the latter being organized into
the classical linear $\real$-xenomorph $\copos$.  In Section 3 we
briefly present how causality, modelled by a partial order between
events, also `uniquely determines' $\cem$, hence infer that a classical
researcher, by using her Boolean logic, builds a picture (model) of the
chronological connection between her event-perceptions effectively
isomorphic to a partial order, with the latter justly called `classical
causality'. In the last Section 4 we borrow $\supra$ from (Raptis,
1998) and organize the non-classical linear $\com$-xenomorph $\qopos$.
After comparing it with $\copos$ we suggest that it is what a `quantal
researcher' perceives as $\qem$ having a quantum version of the
classical causality relation of $\cem$ encoded in the multiplicative
structure of its $\supra$-objects. This is a non-classical linear
$\com$-xenomorph modelling the quantum causal space $\qem$. We find
that a quantal subobject classifier in the quantum topos $\qopos$ is
the Lorentz-spin algebra $sl(2,\com)$ that corresponds to the
relativistic invariances of $\qem$ and is the quantum substitute for
the usual Boolean binary alternative ${\bf 2}$-the subobject classifier
in the classical topos $\copos$ model of $\cem$.  At the end we give
some heuristic arguments to support that this $\qopos$ picture of
$\qem$ agrees with the algebraically quantized causal sets presented in
(Raptis, 1999). In the Conclusion we give a r\'{e}sum\'{e} of the paper
and comment briefly on a possible application of our quantum topos idea
to the problem of quantum gravity.

\section{THE CLASSICAL LINEAR XENOMORPH}

In this section we resume some basic features of classical Linear
Xenology from (Trifonov, 1995).

a. The elementary actions of a researcher effecting events upon the
world form a semigroup with a two-sided identity; a monoid $M$. $M$ is
called the motor or effector space of the researcher. $M$ models our
primitive intuition that the series composition of two classical
(selective) actions is such an action and that this composition is
associative.

b. The observations of a researcher of the events of the world form a
linear space $S$ over a field $\cal{F}$; a vector space $S$. $S$ is
called the sensor or reflexor space of the researcher. $S$ models our
primitive intuition that the classical sense-pictures of the world that
a researcher perceives participate in incoherent, $\cal{F}$-weighed
superpositions and that the latter too are the researcher's mixed,
`probability-weighed' sensors of the world.

c. The Quantum Principle, by positing that every sensor of a researcher
is one of her motors, combines $M$ and $S$ into an associative algebra
$A$ over $\cal{F}$ with an identity. Trifonov follows the
constructivists' jargon and coins the elements of $A$ the researcher's
states of knowledge of the world, while $A$ as a whole, the paradigm of
(the structure of) the world constructed by the researcher.

d. The time proper to the researcher is defined in a constructivistic
sense according to which it partially orders her states of knowledge.
Thus, for a given paradigm $A$ over an ordered field $\cal{F}$ such as
$\real$, time can be represented by a real $1$-form $\tau$ on $A$'s
sensor space $S$, that is, an $\real$-valued linear functional on $A$
as a vector space.

e. There is a `naturally' defined metric $g_{\mu\nu}$ for any given
paradigm $A$ once the proper time functional $\tau$ is fixed on it.
$g_{\mu\nu}$ is induced by $A$'s structure constant $3$-tensor $C(a;\,
b,\, c)$ which is a trilinear functional on two vector arguments $b$,
$c$ in $S$ and on one covector argument $a$ in
$S^{\star}$\footnote{$S$'s linear dual space.} with coordinates
$C^{\lambda}_{\mu\nu}$ in $\cal{F}$. Once $a$ is identified with $\tau$
in $C$ and one insists that the latter be symmetric in the other two
vector arguments, one is left with an $\cal{F}$-valued symmetric tensor
$g_{\mu\nu}$ on $S\times S$.

f. The organization of isomorphic copies of $A$ into a category $\{
A\}$ of linear algebras over $\cal{F}$ having the properties of a topos
is called an $\cal{F}$-xenomorph. An $\cal{F}$-xenomorph may be thought
of as a universe of physico-mathematical discourse members of which,
the researchers, based on their `research parameters of reasoning and
measurement', that is, their logic and $\cal{F}$-coordinatization,
construct a model of the world according to their interactions with
(research on) it.

g. An $\cal{F}$-xenomorph is of (finite) dimension $n$ when its
paradigms are $n$-dimensional vector spaces. If the motor space $M$ of
$A$ is (not) a monoid, $A$ is said to be (ir)rational. An
$\cal{F}$-xenomorph of rational paradigms which is (not) Boolean is
called (non)-classical. When a classical $\cal{F}$-xenomorph consists
of paradigms of finite dimensionality, it is called classic.

h. The basic result from (Trifonov, 1995) is that when the research
parameters are fixed to their current `values', the latter being
Boolean logic for the researcher and $\real$-coefficient field for her
psychology, the only classic paradigm $A$ `compatible' with them is the
$4$-dimensional associative division algebra of real quaternions
$\quat$.

By fixing $\tau$ in $\quat$'s structure constant tensor $C$ and
symmetrizing with respect to the other two vector arguments in it as
explained in e, one extracts the Minkowski metric
$\eta_{\mu\nu}=diag(+1,-1,-1,-1)$ of Lorentzian signature
$|tr(\eta)|=2$ directly from the multiplication table of the unit
quaternions. Thus the motor structure of the classic paradigm $\quat$
determines the special relativistic chronometric $\eta_{\mu\nu}$ on
$S_{\quat}$ and the latter is then identified with Minkowski space
$\cem$. Moreover, once a classical researcher fixes her
(non-relativistic) proper time $\tau$ in $C_{\quat}$, her motor
structure induces in turn the special relativistic chronometric
delimiting the invariant causal structure on her sensor space (the
Minkowski lightcone). Thus proper time, in the constructivistic sense
of a partial order on the researcher's states of knowledge (d), which
is a non-physical mental attribute, generates via the motor structure
of the classical researcher's paradigm $\quat$ the physical metric on
$\cem$.

The Lorentz transformations of special relativity arise naturally from
the categorical arrows between the quaternion paradigms in the topos
$\copos$ that, by their definition as linear homomorphisms, preserve
the linear (sensory) and multiplicative (motor) structure of the
$\quat$ objects in $\copos$ (ie they preserve $\cem$ as a vector space
and its lightcone induced by $\quat$'s multiplicative structure). We
wrap-up this to the following: a Boolean researcher with real proper
time as a partial order on her states of knowledge of the world
uniquely describes the latter as $\cem$ with its special relativistic
causal structure.

i. The concluding remark from (Trifonov, 1995) that we mention here is
that a non-trivial Grassmannian or supersymmetric paradigm
$\cal{G}$, one having non-trivial divisors of zero in its motor
space $M$, is proper to a non-Boolean researcher and her topos $\qset$
is non-classical. Finkelstein (1996) and Selesnick (1998) have
suggested $Qet$, a certain version of the category $\qset$ of Grassmann
algebras, as a quantum replacement of the classical topos $Set$ of
sets.  $Qet$ models quantum set theory as $Set$ is the universe of
classical set theory and the basic insight of Finkelstein is that since
classical spacetime is primarily a classical set soundly represented in
the Boolean topos $Set$, quantum spacetime must first of all be a
quantum set living in the non-classical or `quantal' linear xenomorph
$\qset$\footnote{Finkelstein's original intuition that quantum set
theory should be modelled in the world $Qet$ or $\qset$ appeared in the
literature as early as (1972a, b) and it is all resumed in Finkelstein
(1996).  Significant allusions to the quest for a quantum topos
modelling quantum spacetime can also be found in (Selesnick, 1991,
1998) which is along the lines of Finkelstein's work. To this end
Selesnick too anticipates a quantum replacement of the topos $Set$ by
$Qet$ or some other structure in order to accommodate quantum
spacetime.}. The work of Trifonov (1995) made the subtle contribution
that relativistic spacetime $\cem$ is indeed a conception of a
classical researcher, a classical entity living in a Boolean topos;
however, not in $Set$, but in $\copos$. Of course, if we neglect the
algebraic structure from the objects of $\copos$, from which the metric
structure of $\cem$ derives as described in e and h above, there is a
natural equivalence (functor) between it and $Set$. Since Finkelstein
foremost intended $Qet$ as a universe of discourse in quantum spacetime
topology and not in quantum spacetime metric structure, it serves as a
sound model for the former not the latter\footnote{Afterall, extra
structure is usually imposed on a topological space, such as the
Euclidean manifold $\cem$ one, to become a metric space. $\copos$,
stripped-off its algebraic structure which encodes $\eta_{\mu\nu}$,
becomes a topos of classical sets the Boolean algebra of which suffices
for defining a classical topological space-`a collection of open sets
such that...'.}. For the latter we expect a quantum replacement of
$\copos$ (not $Set$) to serve us as a model of $\qem$: a quantum
Minkowski spacetime.

We provide such a candidate quantum topos in Section 4. First we relate
the $\copos$ picture of $\cem$ above to its causal set one of Sorkin
{\it et al.} (1987).

\section{$\cem$ AS A CLASSICAL CAUSAL SPACE}

In d of Section 2 we mentioned Trifonov's assumption of a
constructivistic notion of time according to which it partially orders
the researcher's states of knowledge of the world. Thus time is not a
physical property of the world {\it per se}, a connection between its
physical events, but, rather indirectly, it is one between the
researcher's perceptions of the events of the world.  That this
psychological conception of time is assumed to be a partial order
however, accords with one of the most successful models for $\cem$ that
has been constructed so far: the causal set of Sorkin {\it et al.}
(1987). The causal set pictures spacetime as a locally finite set of
events with the causal `after' relation between them represented by a
partial order (a locally finite poset).  Sorkin and his coworkers hold
that a partial order determines not only the metric-signature, reality
and four-dimensionality of $\cem$, but also its topological and
differential structure; furthermore, due to its local finiteness, they
maintain that it can serve as a discrete substratum underlying
spacetime at short distances of the order of Planck's ($10^{-33}cm$)
where quantum gravitational effects are expected to be significant.

This application in the context of quantum gravity aside, for the
purposes of the present paper we mention the following affinity between
Classical Linear Xenology and Causal Set Theory in the form of a
conjecture:  since $\cem$ is what a Boolean researcher with a real
psychological proper time uniquely perceives as the event-structure of
the world, and since the same spacetime is uniquely determined by
assuming the causal structure of the events of the world to be a poset,
a partial order is a classical conception of causality intimately
related to a Boolean logical perception of the events of the world and
their chronological connection. A causal set is a classical and classic
causal space in the sense of g in Section 2. It follows that a
non-classical researcher, one that uses a quantal sort of logic and the
usual coefficient field $\com$ for the quantum dynamical amplitudes,
will construct a quantum version of $\cem$, call it $\qem$. This will
be effectively a quantum replacement of the classical topos $\copos$
model for $\cem$ with the resulting quantum topos serving as a model of
quantum causal Minkowski space $\qem$. Many noteworthy attempts have
been made in the past at a straighforward quantum substitution of a
classical causal space of which we mention Finkelstein's spacetime code
(1969), superconducting causal nets (1988) and hypercubical quantum
causal network (1996), as well as this author's more recent try
(Raptis, 1999). In the following section, to realize our conjecture
above, we suggest a non-classical topos as a model of the quantum
causal space $\qem$.

\section{A NON-CLASSICAL TOPOS MODEL OF $\qem$}

Below we introduce a new algebraic paradigm $\supra$ and interpret it
as the local spacetime perceptions of a non-classical researcher.  Then
we `globalize' the result by organizing such objects into a
$\com$-xenomorph, a quantum topos $\qopos$, and maintain that it models
$\qem$ as a quantum causal space. Because a full-fledged presentation
of $\supra$ will take much space thus perhaps disorientate our focus,
we concentrate on the formal properties of $\supra$ that directly
compare with those of the classic paradigm $\quat$ for $\cem$ in
Section 2. This will highlight $\qopos$'s quantal properties in
contradistinction to $\copos$'s classical ones. For more on $\supra$
the reader may refer to the analytic treatment of it in (Raptis, 1998).

${\rm a}^{\prime}$. The linear, sensor structure of $\supra$,
$S_{\supra}$, is that of a $4$-dimensional vector space over the
complex field $\com$ which is the usual coefficient field of quantum
mechanics. As in (Trifonov, 1995) the standard quaternion basis
$\{1,i,j,k\}$ is assumed to be the reference frame proper to a
classical researcher, so in $\supra$ the standard basis is $g=\{
I,\imath,\s,\tau\}$\footnote{The vector $\tau$ in $g$ should not be
confused with the proper time $\tau$ in d of Section 2.}. $g$ is
supposed to be the reference frame proper to a quantal researcher.
Unlike the real psychology $\real$ of the quaternion paradigm of a
classical researcher to which her psychological time takes its values,
orders her states of knowledge of the world and generates the local
causal structure of $\cem$ as in e of Section 2, the complex psychology
$\com$ of a quantal researcher must radically change her conception of
the structure of quantum spacetime. For instance, the principal
difference between the sensor structure of $\supra$ and that of $\quat$
is that the linear superpositions over $\com$ in $S_{\supra}$ are
coherent, that is, quantum amplitude $\com$-weighed ones, while those
in $S_{\quat}$ over $\real$ are incoherent, that is, classical
probability $\real$-weighed ones as mentioned in b of Section 2. The
difference between coherent and incoherent superpositions is a
fundamental one that essentially distinguishes quantum from classical
algebraic structures (Finkelstein, 1996).  For this, $\qem$ as a vector
space, whose quantum paradigm is $\supra$, is isomorphic to $\com^{4}$
rather than to $\real^{4}$ which is the linear structure of the
classical $\quat$ paradigm  of $\cem$. Furthermore, there is a
canonical complex model for quantum Minkowski space associated with
$\com^{4}$, namely, the Grassmannian space
${\cal{G}}_{2}(\com^{4})$ of two dimensional subspaces of
$\com^{4}$ (Selesnick, 1991). The Grassmannian (supersymmetric) nature
of the quantal paradigm $\supra$ for $\qem$ is presented in ${\rm
d}^{\prime}$ below.

${\rm b}^{\prime}$. The multiplicative, motor structure of $\supra$,
$M_{\supra}$, can be resumed in the following table $T(\supra)$ showing
the binary product between the elements of $g$\footnote{From now on,
due to their multiplicative serial compositions, we call the basis
elements in $g$ $\supra$'s generators. The generators of $\supra$ are
interpreted as elementary actions proper to the quantal experimenter
who builds $\supra$ as a paradigm-model of the structure of the world.}

\[ T(\supra)=
\begin{tabular}{|l||l|c|c|r|} \hline
$\supra$&$I$&$\imath$&$\s$&$\tau$\\ \hline\hline
$I$&$I$&$\imath$&$-\tau$&$-\s$\\ \hline
$\imath$&$\imath$&$-I$&$-\tau$&$\s$\\ \hline
$\s$&$\s$&$\tau$&$I$&$-\imath$\\ \hline
$\tau$&$\tau$&$-\s$&$\imath$&$-I$\\ \hline
\end{tabular}\,\,\, .
\]

\noindent Direct inspection of the table shows that the product in
$\supra$ is not associative so that its motor structure is not a
semigroup. Also, $I$ is a right but not a left identity, so that
$M_{\supra}$ is {\it a fortiriori} not a monoid.  In view of the proof
of the basic theorem in (Trifonov, 1995) and h in Section 2, since a
quantal researcher employs a non-Boolean, quantal logic and a complex
psychology to order her states of knowledge of the world, the algebraic
paradigm of the structure of the world that she builds based on them is
expected to be not only non-classical, but also irrational.

From a classical perspective, the `irrationality' of quantal
(selective) actions has already been observed by Finkelstein (1996) and
Selesnick (1998) and can be resumed in the following: the series
composition of two proper (selective) actions may not be a proper
(selective) action. This is a principal characteristic of quantal
actions and, as Finkelstein and Selesnick argue, it is equivalent to
non-commuting proper actions, which is also due to their coherent
superpositions mentioned in ${\rm a}^{\prime}$ above, which in turn
make the logic of the researcher quantum\footnote{Quantum logic is a
non-classical, that is, non-Boolean, logic exactly due to the coherent
superpositions of quanta. In its lattice representations, quantum logic
is modelled after non-distributive lattices, while Boolean lattices are
always distributive. Non-distributivity is due to non-commutativity
which is due to coherent superpositions of quantum actions.}.  This
further qualifies $\supra$ as a prime candidate quantal paradigm to
model $\qem$. At the same time however, the product between proper
actions in $\supra$ seems to violate not only their algebraic closure
in it (ie the series product of two actions in $\supra$ is not a proper
action in $\supra$), or even the existence of
inverses\footnote{$\supra$, unlike the classical quaternion paradigm,
can easily be shown to be not a division ring. In (Trifonov, 1995) it
is argued that if the logic of the researcher is Boolean the motor
structure of her proper paradigm of the world is a group. For $\quat$
in particular, $M_{\quat}$ is the multiplicative group of non-zero
quaternions isomorphic to $SU(2)\times\real^{+}$. A non-classical,
quantal logic for the researcher is consistent with non-invertible
motor actions on the world. We return to some implications of this
remark in the Conclusion at the end.}, but also the associative law.
Finkelstein notes and asks in (1996): \begin{quotation}...Our selective
acts for the quantum do not all commute. It follows that the
composition of two selective acts in series is not always a selective
act...On the other hand, the associative law $A(BC)=(AB)C$ seems to
persist. Can you see its empirical meaning ? What experiment would
break it ?  ...Because the concatenation of two selective acts is a
more general kind of action, it is artificial and clumsy to separate
logic from dynamics in quantum theory as we do in classical
thought...\end{quotation} Since the associative motor structure of the
classic paradigm $\quat$ encodes the (local) causal structure of
$\cem$\footnote{Section 2e, also witness in ${\rm c}^{\prime}$ that
follows how the local Lorentzian chronometric $\eta_{\mu\nu}$ is
effectively encoded in the binary product between the generators of
$\quat$.}, we expect the non-associativity of the product in $\supra$
to capture an essential trait of (local) `quantum causality'. In ${\rm
g}^{\prime}$ we give, along this `logical-motor-causal' line of
thought, a plausible answer to Finkelstein's question on the physical
meaning of non-associativity and in the Conclusion we address his `{\it
logics comes from dynamics}'\footnote{From the Foreword `About the
Philosophy' in (Finkelstein, 1996).} in the opposite way: `(quantum)
causality comes from (quantum) logics'.

${\rm c}^{\prime}$. Like in d and e of Section 2, in $\supra$ too we
identify a proper time linear functional $\tau$ on $S_{\supra}$ taking
its values in $\com$ and such that, when identified with $a$ and
evaluated in the structure constant tensor $C(a;\, b,\, c)$ of
$\supra$, it induces a (quantum) spacetime metric proper to the
paradigm. Let us recall in more detail from (Trifonov, 1995) how this
is done.

For the classical paradigm $\quat$ of $\cem$ the components
$g_{\mu\nu}$ of the metric in some basis $\{ e_\mu\}$ and its dual
$\{ e^{\nu}\}$ are given by

\[
\begin{array}{rcl}
&g_{\mu\nu}=C(\tau ;b,c)=C(\tau_{\lambda}e^{\lambda};
b^{\mu}e_{\mu},c^{\nu}e_{\nu})=
\tau_{\lambda}C^{\lambda}_{\mu\nu}\,\,.\cr
\end{array}
\]

\noindent Then, in the standard basis $\{ 1,i,j,k\}$ for $\quat$
$g_{\mu\nu}$ reads

\[
\begin{tabular}{|l|c|c|c|r|} \hline
$\tau_{0}$&$\tau_{1}$&$\tau_{2}$&$\tau_{3}$\\ \hline
$\tau_{1}$&$-\tau_{0}$&$-\tau_{3}$&$\tau_{2}$\\ \hline
$\tau_{2}$&$\tau_{3}$&$-\tau_{0}$&$-\tau_{1}$\\ \hline
$\tau_{3}$&$-\tau_{2}$&$\tau_{1}$&$-\tau_{0}$\\ \hline
\end{tabular}\,\,\,.
\]

\noindent Notice that it effectively corresponds to the product table
$T(\quat)$ of the standard unit quaternions multiplied by the
components $\tau_{\mu}$ of the proper time in the dual standard basis
for $\quat$. To see this just identify $\tau_{0}\equiv 1$,
$\tau_{1}\equiv i$, $\tau_{2}\equiv j$ and $\tau_{3}\equiv k$ in the
table above and get the entries in $T(\quat)$ below

\[ T(\quat)=
\begin{tabular}{|l||l|c|c|c|} \hline
$\quat$&$1$&$i$&$j$&$k$\\ \hline\hline
$1$&$1$&$i$&$j$&$k$\\ \hline
$i$&$i$&$-1$&$-k$&$j$\\ \hline
$j$&$j$&$k$&$-1$&$-i$\\ \hline
$k$&$k$&$-j$&$i$&$-1$\\ \hline
\end{tabular}\,\, {\rm with}\,\, kji=-1.
\]

\noindent By insisting that $g_{\mu\nu}$ is a symmetric matrix we get
$\tau_{1}=\tau_{2}=\tau_{3}=0$ and we recover, up to a scalar factor
$\tau_{0}$, the Lorentz metric $\eta_{\mu\nu}=diag(1,-1,-1,-1)$ with
trace of absolute value $2$.  In this sense `{\it if we ignore the
motor structure of the paradigm, four-dimensionality and Lorentz metric
become a mystery}' (Trifonov, 1995).  The Lorentz metric is effectively
encoded in the binary multiplication table of the unit quaternions.
Also notice that $\eta_{\mu\nu}$ is the component $C^{0}_{\mu\mu}$ of
the structure constant tensor $C(\quat)$ and it is generated by the
`wrist-watch' psychological time $(\tau_{0}, 0,0,0)$. In the product
table of the quaternions this is the $\pm 1$ entries along the main
diagonal corresponding to the squares of the four unit quaternions.

Similarly, for the quantal paradigm $\supra$ of $\qem$ the components
of the metric in the standard basis $g$ are derived as for
$\eta_{\mu\nu}$ in $\quat$ above so that, after symmetrization and
along $\tau_{0}$, we read from $T(\supra)$ the diagonal metric matrix
$\kappa_{\mu\nu}=diag(1,-1,1,-1)$ of signature $0$. This is the
traceless Klein metric. The physical significance of this form of the
metric for $\qem$ will be given shortly when we discuss the special
relativistic and causal symmetries of $\supra$ associated with
$\kappa_{\mu\nu}$.  First we have to show the Grassmannian or
supersymmetric character of $\supra$ as promised at the end of ${\rm
a}^{\prime}$.

${\rm d}^{\prime}$. The sensor space $S_{\supra}$ is $\aker_{2}$-graded
as follows

\[
\begin{array}{rcl}
&S_{\supra}=S_{\supra}^{0}\oplus S_{\supra}^{1}=
\spn_{\com}\{ I,\imath\}\oplus span_{\com}\{\s ,\tau\}\cr
\end{array}
\]

\noindent with $S_{\supra}^{0}$, spanned by $I$ and $\imath$, the even,
complex, $2$-dimensional subspace of $\com^{4}$ of grade $0$ elements
of $\supra$ and $S_{\supra}^{1}$, spanned by $\s$ and $\tau$, the odd,
complex, $2$-dimensional subspace of $\com^{4}$ of grade $1$ elements
of $\supra$. This makes $\supra$ a Lie superalgebra (Raptis, 1998),
thus it is a Grassmannian or supersymmetric paradigm in the sense of i
of Section 2. The $(2,2)$ split of $S_{\supra}\simeq\com^{4}$ by grade
effects a natural isomorphism between $\supra$ and the Grassmannian
paradigm ${\cal{G}}_{2}(\com^{4})$ of two dimensional subspaces of
complex Minkowski space used by Selesnick (1991) to model $\qem$ and
alluded to at the end of ${\rm a}^{\prime}$.  Thus $\supra$ qualifies
even further as a sound model of $\qem$.

The physical interpretation of members of $S_{\supra}^{0}$ is bosons
and of those in $S_{\supra}^{1}$ fermions. We may define the parity (or
grade) binary alternative $\pi(x)$ by

\[
\begin{array}{rcl}
&\pi(x):=
\left\lbrace \begin{array}{rcl}
0& \mbox{if}\,\, x\in S_{\supra}^{0}, \cr
1& \mbox{if}\,\, x\in S_{\supra}^{1}, \cr
\end{array}\right. \cr
\end{array}
\]

\noindent and tentatively interpret the direct sum ($\oplus$) $(2,2)$
split of $S_{\supra}$ above as the usual SUSY Wick-Wightman-Wigner
spin-statistics superselection rule that forbids the coherent quantum
superpositions of bosons with fermions (Freund, 1986)\footnote{Some
authors like Freund (1986) write $V=V^{0}\cup V^{1}$ for this
incoherent $\aker_{2}$ split by grade of a vector space $V$ into its
even and odd subspaces, others like Okubo (1997) write $V=V^{0}\oplus
V^{1}$ as we do.}. However, due to our earlier interpretation in ${\rm
a}^{\prime}$ of $+$ as coherent quantum superposition throughout all
$S_{\supra}$, we anticipate a lifting of the spin-statistics
superselection rule with concomitant formal replacement of the
incoherent direct sum $\oplus$ by coherent quantum superposition $+$.
In $\supra$ bosons are allowed to superpose with fermions\footnote{The
physical consequences for the quantum structure of spacetime of such a
lifting of the spin-statistics superselection rule are currently under
investigation by this author. For instance, since the spin-statistics
connection and the spin-statistics superselection rule are `theorems'
that derive from the `causal axiom' of Einstein Locality in quantum
field theory on $\cem$ (Haag, 1992), a quantum causal theory for $\qem$
perhaps makes the latter a non-valid assumption so that the
spin-statistics superselection rule is violated.}.

Since $\supra$ is a Grassmannian paradigm it has non-trivial divisors
of zero\footnote{For example, $\tau+\s\in S_{\supra}^{1}$ and
$(\tau+\s)^{2}=0$.}, thus $M_{\supra}$, unlike $M_{\quat}\simeq
SU(2)\times\real^{+}$, is not a group and its logic is non-Boolean
(Trifonov, 1995; see also footnote 11 in ${\rm b}^{\prime}$). We
emphasize again, $\supra$ is a paradigm for $\qem$ of a quantal
researcher.

${\rm e}^{\prime}$. We organize the quantal paradigms $\supra$ into the
non-classical $\com$-xenomo\- rph $\qopos$. $\qopos$ qualifies as a
quantum topos model of $\qem$ (Raptis, 1998, Selesnick, 1998). A major
difference in structure between the quantum topos $\qopos$ for $\qem$
and its classical counterpart $\copos$ for $\cem$ is one concerning
their subobject classifiers $\Omega$. $\copos$, being a classical
topos, has $\Omega={\bf 2}=\{0,1\}$, the set of Boolean truth values,
as subobject classifier, while $\qopos$, being $\copos$'s quantal
analogue, is expected to have a quantum version of ${\bf 2}$ as
subobject classifier.  Selesnick (1994, 1998), working on Finkelstein's
quantum set theory, net dynamics and relativity, found that a plausible
candidate for such a quantum version of the classical binary
alternative ${\bf 2}$ of the Boolean topos $Set$ is $SL(2,\com)$, the
Lorentz-spin group of special relativity.

For a technical definition of the subobject classifier the reader is
referred to (Goldblatt, 1984)\footnote{For an excellent treatment of
categories and topoi from a (classical) logical point of view the
reader is also referred to (Lambek and Scott, 1986). A good treatment
of topoi from a local set theoretic viewpoint is (Bell, 1988).}.
Rougly, and deriving from the classical topos $Set$ of sets, the
subobject classifier, tells one whether an object in the topos is
`smaller than' (included in, injected into, a subobject of) another or
not. Thus its values are in some sense the `inclusion symmetries' of
the objects in the topos, as for example in $Set$, whether a set is
included in another ($\Omega=1$) or not ($\Omega=0$).  A familiar use
of ${\bf 2}$ in $Set$ is that for every set $\alpha$ its power set
${\bf 2}^{\alpha}$ is the one containing all the subobjects (subsets)
of $\alpha$. Then, in a subtle sense, the quantal binary symmetries
$SL(2,\com)$ of quantum spacetime modelled after the world
$Qet\simeq\qset$ of quantum sets, is the invariance group of the causal
structure of special relativistic spacetime $\cem$ as perceived by a
macroscopic, classical observer (Selesnick, 1994, 1998). The analogy
between $Set$ and $Qet$ in this respect is the following: the inclusion
$\beta\in\alpha$ or the injection $\beta\rightarrow\alpha$ between
classical sets in $Set$ practically corresponds to the causal
connection $\beta\rightarrow\alpha$ between quantum spacetime events in
$Qet$, so that $SL(2,\com)$, like ${\bf 2}$ in $Set$, is the invariance
group of the past lightcone
$\lambda(\alpha):=\{\beta:\,\,\beta\rightarrow\alpha\}$ of a quantum
spacetime event $\alpha$ in $Qet$\footnote{For instance, for the
quantum causal nets in (Finkelstein, 1988), $SL(2,\com)$ is held to
correspond precisely to the symmetry group acting on the two spinor
inputs of the binary quantum causal cell.}.

Still though, as Selesnick (1994, 1998) emphasizes, the group
$SL(2,\com)$, the coherent state exponential of the Lie algebra
$sl(2,\com)$ of the quantum binary alternative, is what a coarse,
macroscopic, classical observer with a limited power of resolution
perceives as the invariance group of the underlying quantum Minkowski
plenum $\qem$. Our quantum topos $\qopos$ by comparison models the
states of knowledge of $\qem$ of a fine, `gedanken microscopic',
quantal researcher who is supposed to possess higher power of
resolution than her classical counterpart\footnote{For instance, as we
have seen she is ideally assumed to perceive coherent quantum
superpositions in her sensory space $S$, while her classical
counterpart is limited to `sense' only incoherent superpositions.},
thus operate at the quantum (algebra) not the classical (coherent
state, group) level. Below we suggest an entirely discrete-algebraic
(combinatorial) scheme, solely at the algebra level of $\supra$, that
leads directly  to the Lorentz-spin structure $sl(2,\com)$, which then
serves as the quantal subobject classifier in the quantum topos
$\qopos$ model of $\qem$. This scheme is mainly based on quantum topos
ideas that first appeared and were analytically treated in (Raptis,
1998).

${\rm f}^{\prime}$. To derive the Lorentz-spin algebra $sl(2,\com)$ as
the causal symmetries of the quantum topos $\qopos$ for $\qem$, we look
at the two $(2,2)$ splits of $S_{\quat}$ by norm in ${\rm c}^{\prime}$
and parity in ${\rm d}^{\prime}$. The first pertains to the $\mu=\pm1$
signature binary distinction of the four generators of $\supra$ in the
Klein metric $\kappa_{\mu\nu}=diag(+1,-1,+1,-1)$, while the second to
their $(-1)^{\pi(x)}=\pm1$ grade binary distinction which we still call
$\pi$. These two binary characteristics of the four generators of
$\supra$ are resumed in the following table

\[
\begin{tabular}{|c||c|r|} \hline
${\bf 2}\times{\bf 2}$&$\mu$&$\pi$\\ \hline\hline
$I$&$+1$&$+1$\\ \hline
$\imath$&$-1$&$+1$\\ \hline
$\s$&$+1$&$-1$\\ \hline
$\tau$&$-1$&$-1$\\ \hline
\end{tabular}\,\,\,.
\]

\noindent Then we assume that the Klein four-group ${\bf 4}_{2}={\bf
2}\times{\bf 2}$ is the permutation symmetry group of the four
generators of $\supra$. By indexing the latter in $g$ as $I=g_{0}$,
$\imath=g_{1}$, $\sigma=g_{2}$ and $\tau=g_{3}$, the Klein four-group
${\bf 4}_{2}=\{ (0)(1)(2)(3), (01)(23), (03)(14), (04)(13)\}$ permutes
these indices and transposes their corresponding generators, of norm
($\mu$) and parity ($\pi$) characteristics as in the table above, as
follows

\[
\begin{array}{rcl}
&(0)(1)(2)(3)=\{ I,\imath ,\s ,\tau\}=
 {\rm `no\,\,permutation'\,\, (identity)},\cr
&(01)(23)=\{ I\leftrightarrow \imath ,\s\leftrightarrow\tau\}=
 {\rm `norm\,\, swap\,\, at\,\, constant\,\, parity'},\cr
&(02)(13)=\{ I\leftrightarrow\s ,\imath\leftrightarrow\tau\}=
 {\rm `parity\,\, swap\,\, at\,\, constant\,\, norm'},\cr
&(03)(12)=\{ I\leftrightarrow\tau ,\imath\leftrightarrow\s\}=
 {\rm `norm\,\, and\,\, parity\,\, swap'}.
\end{array}
\]

\noindent One may think of the norm and parity binary values of the
$g_{\mu}$s as their `causal colors' and the ${\bf 4}_{2}$ acting on
them as their `causal color group'.  Straightforward quantization of
${\bf 4}_{2}$ \`{a}-la Finkelstein (1996) or Selesnick (1998) yields
the Lorentz-spin algebra $sl(2,\com)$, the `quantum causal color
symmetry structure', which a fine, quantal researcher perceives as the
relativistic, causal symmetries of the quantum topos $\qopos$ model of
$\qem$-its quantal subobject classifier. The whole dynamical or causal
algebraic scheme in the quantum topos $\qopos$ may then be called
`quantum causal chromodynamics'.  From the same works (Finkelstein,
1996, Selesnick, 1998) it follows that a coarse, macroscopic, classical
observer perceives coherent exponentials of $sl(2,\com)$ which
correspond to elements of the group $SL(2,\com)$-the quantal version of
the classical subobject classifier ${\bf 2}$ of the Boolean topos
$\copos$ model of $\cem$.

In closing ${\rm f}^{\prime}$ we mention an affinity between our
$\qopos$ model of $\qem$ as a (quantal) causal space and Finklestein's
null tesseractal (quantum) causal net ${}_{q}\fuss^{4}$ (Finkelstein,
1996).  The discrete causal symmetries of the latter constitute the
symmetric group $S_{4}$ of $4!=24$ permutations that act on a frame $\{
n_{\mu}\}$ consisting of four linearly independent null vectors (null
frame) and leave the following symmetric null form $n_{\mu\nu}$  of
their mutual inner products invariant

\[ n_{\mu\nu}=\left(
\begin{tabular}{lccr}
$0$&$1$&$1$&$1$\\
$1$&$0$&$1$&$1$\\
$1$&$1$&$0$&$1$\\
$1$&$1$&$1$&$0$\\
\end{tabular}\,\,\,
\right)\,\, .
\]

\noindent $S_{4}$ can be semi-factorized as ${\bf 4}_{2}\oslash{\bf
3}\oslash{\bf 2}\oslash{\bf 1}$ with ${\bf 4}_{2}$ the discrete version
of the Lorentz-spin spacetime group as above, ${\bf 3}$ the discrete
ancestor of the GUT color gauge group $SU(3)$ and ${\bf 2}\oslash{\bf
1}$ the discrete precursor of the GUT electroweak gauge group
$SU(2)\times U(1)$.  The $S_{4}$-symmetric null frame and form for
$\cem$ is preferred over the Minkowskian $1+3$ ($1+\{ i,j,k\}$)
quaternion frame and Lorentz metric $\eta_{\mu\nu}$ having only
$S_{3}$, the six permutations of its $3$ unphysical spacelike vectors,
as causal symmetries. Thus, even the internal, gauge symmetries are
accounted for in an `external', causal way, something that is not
possible in the quaternionic, Lorentzian $1+3$ picture of $\cem$.

In $\supra$ on the other hand, $\kappa_{\mu\nu}$ is traceless like
$n_{\mu\nu}$ and ${\bf 4}_{2}$ also stands for the discrete,
classically thought of as external, spacetime symmetries of $\cem$,
albeit, of a complexified version of it ($\com^{4}$).  We found that in
$\supra$ even these external spacetime symmetries arise from permuting
the `(quantum) causal colors' of its generators.  The latter are
characteristic physical properties of $\supra$'s generators that are
intimately related to quantum integral/half-integral spin (parity
$\pi=+1/-1$) and to relativistic spacelike/timelike (norm $\mu=+1/-1$)
binary distinctions, respectively. Could it be that spacetime itself
arises from some deep inner quantum relativistic algebraic distinctions
such as these ?\footnote{For a deep treatment of the notion of
`distinction' and how it may give rise to algebraic (spacetime)
structures the reader is referred to (Kauffman, 1991).} If yes, how is
the relativistic `spacelike' ($\mu=+1$) character related to the
quantum `bosonic' ($\pi=+1$) one and the timelike ($\mu=-1$) to the
fermionic ($\pi=-1$) ? Furthermore, while in (Raptis, 1998) an attempt
to answer to these questions is made, it is also shown that one is able
to form a null-frame like $\{ n_{\mu}\}$ solely from algebraic
associations of the generators of $\supra$ and then straightforwardly
apply Finkelstein's $S_{4}$ ideas above.

${\rm g}^{\prime}$. From ${\rm a}^{\prime}$ to ${\rm f}^{\prime}$ we
arrived at the quantum topos $\qopos$ as a cogent model of the quantum
causal space $\qem$ having as quantum causal symmetry structure the
quantal subobject classifier $sl(2,\com)$. The latter's coherent form
$SL(2,\com)$, the double cover of the Lorentz group, is the quantum
version of the Boolean binary alternative ${\bf 2}$-the subobject
classifier of the classic xenomorph $\copos$. The classical researcher
building the $\copos$ model of $\cem$ perceives $SL(2,\com)$ in a
`sensory-motor' way (Trifonov, 1995) as the linear isometries relating
various quaternionic frames $e_{\mu}$ of $\quat$. It follows from
Section 3 that while $\copos$ is a model of classical causal space
$\cem$ equivalent to a partial order between events and coined
`classical causality', $\qopos$ is a model of quantum causal space
$\qem$ with an order relation between events other than a partial
order.

The inadequacy of a partial order to model `quantum causality' has been
noticed by Finkelstein (1988, 1996) and recently exposed by this author
(Raptis, 1999) in an attempt to algebraically quantize Sorkin {\it et
al.}'s causal sets (1987).  It was also explicitly anticipated in
(Raptis, 1998).  Briefly, a partial order fails to model `quantum
causality' on grounds of locality: because a partial order is
transitive, thus mediated, it is not local, when in fact we suppose
that all the fundamental (quantum) variables in Nature are local.
Hence, we posit that `quantum causality' is an intransitive, thus
immediate, relation between events that should be algebraically
represented. Such a model of `quantum causality' was presented in
(Raptis, 1999). Here we also propose an algebraic model\footnote{A
$\underline{\rm local}$ algebraic model since the paradigm $\supra$ is
a (super)-Lie algebra (Raptis, 1998).}, a paradigm of $\qem$ as a
quantum causal space, thus it is natural to ask which algebraic feature
of $\supra$ corresponds to the desirable intransitivity or immediacy of
`quantum causality' that it aspires to encode in its motor structure.

We pick the argument from the end of ${\rm b}^{\prime}$ and suggest
that it is the non-associativity of the binary product in $\supra$ that
corresponds to the intransitivity, for locality's sake, of the quantum
causal connection in $\qem$ that $\supra$ algebraically models. We show
this by the following heuristic argument which, while non-rigorous, is
rather intuitive and ostensive: let $ba$, the series concatenation of
two generators in $\supra$ from right to left, or equivalently, the
series composition of two actions of the researcher in her motor space,
stand for an immediate causal link $b\leftarrow a$. Then, the
(in)transitivity of `$\leftarrow$' can be symbolically, algebraically
and equationally cast as (non)associativity

\[
\begin{array}{rcl}
&[(c\leftarrow b)\wedge(b\leftarrow a)\Rightarrow
 (c\leftarrow a)]\Leftrightarrow [(cb)a=c(ba)]\,\, ,\cr
&[(c\leftarrow b)\wedge(b\leftarrow a)\not\Rightarrow
 (c\leftarrow a)]\Leftrightarrow [(cb)a\not=c(ba)]\,\, .
\end{array}
\]

\noindent It follows that the classical, associative quaternion
paradigm $\quat$ is a sound algebraic model of the `classical causal
space' $\cem$ which is supported by a classical causal relation: the
transitive, hence non-local (mediated), partial order. On the other
hand, the quantal non-associative paradigm $\supra$ is a sound
algebraic model of the `quantum causal space' $\qem$ which is supported
by a quantal causal relation: the intransitive, hence local
(immediate), succession relation (Finkelstein, 1996).

\section{CONCLUSION CUM DISCUSSION}

The present paper follows Trifonov's project (1995), to model the
measurement (active observation) process and express how the logic of a
classical observer determines what she `sees', and extends it for the
`active observation parameters' of a quantal researcher. The latter, we
argued, builds the quantum topos $\qopos$ picture of the causal
structure of quantum Minkowski spacetime $\qem$. The subobject
classifier of this non-classical linear xenomorph is an algebraic,
quantum version of the Boolean binary alternative of the classical
topos $\copos$ model of classical Minkowski space $\cem$ and
corresponds to the Lorentz-spin algebra $sl(2,\com)$.  The resulting
algebraic picture of $\qem$ as a quantum causal space was seen to agree
in some manner with the recently algebraically quantized causal sets of
Sorkin {\it et al.} (Raptis, 1999).

In a way our approach is opposite to Finkelstein's `(quantum) logics
come from (quantum) dynamics' (1996) and may be squeezed to the motto
`(quantum) causality comes from (quantum) logic'. Truth must lie at a
synthesis of these two dual conceptions so that, as Finkelstein also
points out in (1996), it is quite unnatural to separate logics from
dynamics and causal structure in the quantum deep\footnote{For such a
synthesis see end of quotation from (Finkelstein, 1996) in Section
4${\rm b}^{\prime}$. In view of the present paper, such a synthesis may
be called `quantum causalization of the observer' and it pertains to
regarding the researcher as an indiscriminable part of the deep quantum
dynamical process.}.

We conclude this paper with a future project on an open problem. Our
approach to quantum spacetime's causal structure via the quantum topos
$\qopos$ is by no means complete. $\qopos$ models quantum Minkowski
space that simply lacks gravity: it is globally flat.  The `true'
quantum topos for spacetime should be able to accommodate some sort of
`quantum gravity'. Since gravity may be thought of as the dynamics of a
variable causal connection, the force that tilts the lighcone at every
event, and since a topos can be viewed as a realm of (dynamically)
variable entities\footnote{The topos $Set$ for instance can be viewed
as a world of varying sets (Bell, 1988).}, so the quantum topos
$\qopos$ may be thought of as being only locally $\qem$, that is, be
naturally `gauged' to a twisted or curved sheaf of local paradigms
$\supra$. Thus it may eventually provide a theoretical scheme for
describing the dynamics of the quantal lightcone (quantum causality)
encoded locally in each of its stalks $\supra$.  As we saw in the
present paper, this encodement is in the motor structure of each of its
stalks $\supra$, hence an abstract algebraic quantum topos scenario for
quantum gravity may be expressed thus: `quantum gravity is a theory of
the dynamics of an algebraic product in a quantum topos/sheaf of
locally flat $\qem$-algebras'\footnote{Since we also saw in footnote 11
of 4${\rm b}^{\prime}$ that in $\supra$ there are no inverses, it also
has a good chance of algebraically representing the microlocal `quantum
arrow of time' that is supposed to be the key feature of the `true
quantum gravity' (Penrose, 1987).  The quantum causal nets in
(Finkelstein, 1988) are time-asymmetric too, being described by chiral
spinors. Also Selesnick, working on Finkelstein's quantum net, has
noticed a similar asymmetry in that spacetime quanta are solely
`left-handed' by transforming exclusively under the regular
representation of $SL(2,\com)$ and not its conjugate which represents
`right-handed' quanta (Selesnick in private correspondence). Finally,
in (Raptis, 1998) $\supra$ is seen to be `multiplicatively directed',
that is, its generators in $g$ participate in serial products only in a
certain order resembling the normal or time-ordered products of
operators in quantum field theory. On top of this asymmetry, very
strong gravitational forces, such as those expected at Planck scale,
should significantly tilt the lightcone so as to render causality an
intransitive relation (Auyang, 1995); hence, the non-associative
product in $\supra$ (Section 4${\rm g}^{\prime}$) is all the more
qualified to algebraically represent such a `time-asymmetric curved
causality' (Raptis, 1998).}. The crucial question seems to be: how do
we vary an algebraic product, or what amounts to the same, what is the
curvature of an algebraic product?

\vskip 0.2in

\noindent {\bf\large ACKNOWLEDGMENTS}

\vskip 0.15in

Wholehearted thanks to Professors Selesnick and Lambek for their
incessant support and encouragement during this author's quest for `the
true quantum topos of Nature'. Exchanges on quantum causality with
Professor Mallios are also kindly acknowledged.

\vskip 0.2in

\noindent {\bf\large REFERENCES}

\vskip 0.15in

\noindent Alexandroff, A. (1956). {\it Helvetica Physica Acta}, {\bf 4}
(Suppl.), 44.

\noindent Auyang, S. Y. (1995). {\it How is Quantum Field Theory
Possible ?}, Oxford University Press, Oxford.

\noindent Bell, J. L. (1988). {\it Toposes and Local Set Theories},
Oxford University Press, Oxford.

\noindent Bombelli L., Lee J., Meyer D. and Sorkin R. D. (1987). {\it
Physical Review Letters}, {\bf 59}, 521.

\noindent Finkelstein, D. (1969). {\it Physical Review}, {\bf 184},
1261.

\noindent Finkelstein, D. (1972a). {\it Physical Review}, {\bf D5},
320.

\noindent Finkelstein, D. (1972b). {\it Physical Review}, {\bf D5},
2922.

\noindent Finkelstein, D. (1988). {\it International Journal of
Theoretical Physics}, {\bf 27}, 473.

\noindent Finkelstein, D. R. (1996). {\it Quantum Relativity: A
Synthesis of the Ideas of Einstein and Heisenberg}, Springer-Verlag,
Berlin-Heidelberg-New York.

\noindent Freund, P. G. O. (1989). {\it Introduction to Supersymmetry},
Cambridge University Press, Cambridge.

\noindent Goldblatt, R. (1984). {\it Topoi: The Categorial Analysis of
Logic}, North-Holland, Amsterdam.

\noindent Haag, R. (1992). {\it Local Quantum Physics},
Springer-Verlag, Berlin-Heidelberg-New York.

\noindent Kauffman, L. (1991). {\it Knots and Physics}, World
Scientific, Singapore-New Jersey-London-Hong Kong..

\noindent Lambek, J. and Scott, P. J. (1986). {\it Introduction to
Higher Order Categorical Logic}, Cambridge University Press, Cambridge.

\noindent Okubo, S. and Kamiya N. (1997). {\it Journal of Algebra},
{\bf 198}, 388.

\noindent Penrose, R. (1987). {\it 300 Years of Gravitation}, Eds.
Hawking, S.W. and Israel, W., Cambridge University Press, Cambridge.

\noindent Raptis, I. (1998). {\it Axiomatic Quantum Timespace
Structure: A Preamble to the Quantum Topos Conception of the
(Minkowski) Vacuum}, Ph.D. Thesis, University of Newcastle upon Tyne.

\noindent Raptis, I. (1999). {\it An Algebraic Quantization of Causal
Sets}, submitted to the International Journal of Theoretical Physics
and electronically posted at gr-qc/9906103.

\noindent Robb, A. A. (1914). {\it Theory of Time and Space}, Cambridge
University Press, Cambridge.

\noindent Robb, A. A. (1921). {\it The Absolute Relations of Space and
Time}, Cambridge University Press, Cambridge.

\noindent Selesnick, S. A. (1991). {\it International Journal of
Theoretical Physics}, {\bf 30}, 1273.

\noindent Selesnick, S. A. (1994). {\it Journal of Mathematical
Physics}, {\bf 35}, 3936.

\noindent Selesnick, S. A. (1998). {\it Quanta, Logic and Spacetime:
Variations on Finkelstein's Quantum Relativity}, World Scientific,
Singapore-New Jersey-London-Hong Kong.

\noindent Trifonov, V. (1995). {\it Europhysics Letters}, {\bf 32},
621.

\noindent Zeeman, E. C. (1964). {\it Journal of Mathematical Physics},
{\bf 5}, 490.

\noindent Zeeman, E. C. (1967). {\it Topology}, {\bf 6}, 161.

\end{document}